\documentclass[epj]{svjour}

\usepackage{epsfig}

\usepackage{graphicx,latexsym,xcolor}
\usepackage{amsmath}

\global\long\def\bege{\begin{equation}}
\global\long\def\ende{\end{equation}}

\global\long\def\begal{\begin{align}}
\global\long\def\endal{\end{align}}

\title{Emergent spacetime from purely random structures\\}

\author{Ioannis Kleftogiannis\inst{1} \and Ilias Amanatidis\inst{2} }

\institute{                    
  \inst{1}  Physics Division, National Center for Theoretical Sciences, Hsinchu 30013, Taiwan\\
  \inst{2} Department of Physics, Ben-Gurion University of the Negev, Beer-Sheva 84105, Israel
}

\date{\today \\ \email{ph04917@yahoo.com, eliasamanatidis1@hotmail.com}}

\abstract{
We examine the fundamental question whether a random discrete structure with the minimal number of restrictions can converge to continuous metric space. We study the geometrical properties such as the dimensionality and the curvature emerging out of the connectivity properties of uniform random graphs. In addition we introduce a simple evolution mechanism for the graph by removing one edge per a fundamental quantum of time from an initially complete graph. We show an exponential growth of the radius of the graph, that ends up in a random structure with emergent average spatial dimension $D=3$ and zero curvature $K=0$, resembling a flat 3D manifold, that could describe the observed space in our universe and some of its geometrical properties. In addition, we introduce a generalized action for graphs based on physical quantities on different subgraph structures that helps to recover the well known properties of spacetime as described in general relativity, like time dilation due to gravity. Also, we show how various quantum mechanical concepts such as generalized uncertainty principles based on the statistical fluctuations can emerge from random discrete models. Moreover, our approach leads to a unification of space and matter-energy, for which we propose a mass-energy-space equivalence that leads to a way to transform between empty space and matter-energy via the cosmological constant. 
}

\begin{document}

\maketitle

\section{Introduction}
Spacetime has been assumed for many centuries as a fundamental structure providing a background, acting as a stage for the dynamical evolution of quantities such as matter and energy.
General relativity(GR) has shown that spacetime has to be considered as dynamical also, influenced by the physical quantities contained in it. This approach has had enormous successes in describing various astrophysical and cosmological properties of the observed universe. However unresolved issues like the small value of cosmological constant which cannot be simply interpreted as originating from quantum field fluctuations that fill empty space, strongly imply that space and its geometrical properties like the dimensionality and the curvature should emerge from more fundamental structures. There have been various attempts based on discrete models which aim to eludicate the origin of spacetime and its properties\cite{sunil,dienes,rovelli1,rovelli2,bombelli,fay1,fay2,surya,wolfram,gorard,markopoulou,lombard,trugenberger,trugenberger2,trung,loll1,loll2,klitgaard,klitgaard1}.
Our approach based on random graphs (networks) shows that different generic geometric and structural concepts like the dimensionality, the curvature and an exponential rate of expansion are strongly tied together and can emerge as interlinked intrinsic properties of space.
Unlike other approaches, where the graph/network is usually preembedded in a metric space/manifold, or imposing restrictions in random regular graph approaches, simplicical decomposition of space or triangulation of continuous Riemannian manifolds, our method is based on a purely random discrete structures, uniform random graphs which contain the most minimal number of restrictions. We examine the fundamental question whether such a random discrete structure can converge to a continuous metric space. We find emergent properties that resemble the space observed in our current universe that are fully determined by the connectivity properties of the uniform random graphs, in a background independent manner. In addition our approach leads to a unification of matter-energy and space expressed via a mass-energy-space equivalence. Furthermore we show how quantum mechanical concepts such us uncertainty principles for different physical quantities can arise from the statistical fluctuations of random discrete models.

\section{Graph model}
Consider a simple connected graph $G=(V,E)$ \cite{farkas,newman,frieze,bollob,cohen,kosmidis} where $V(G)=\big\{ v_{1},v_{2},..,v_{n} \big\}$ is a set of n vertices and $E(G)=\big\{e_{1}, e_{2},...,e_{m} \big\}$ a set of m vertex pairs (edges) $e_{i}=\big\{v_{j},v_{k} \big\}$, randomly distributed among the vertices. There are $\Omega=\binom{ \binom{n}{2}}{m}$ possible configurations of the m edges among the n vertices. We consider the simplest case, a uniform random graph where all the configurations have the same probability to appear $P=1/\Omega$. Other useful quantities are, the number of neighbors at each vertex $d(i)$, the so called degree, and the ratio of edges over vertices $R=\frac{m}{n}$, which is double the average degree over all the vertices $\langle d(i) \rangle=2R=\frac{2m}{n}$ (for $n \gg 1$). Since all the possible configurations are equally likely to appear, with the same probability, uniform random graphs have the maximum  entropy, compared to other random structures. If we treat each configuration as a microstate then the Gibbs entropy of the uniform random graph is given by 
\begin{equation}
S_{Gibbs}=ln\Omega=ln\binom{\binom{n}{2}}{m}=ln\binom{\binom{n}{2}}{Rn}.
\label{entropy_gibbs}
\end{equation} 
There are two limits for the above equation.
For $m=0$ and $R=0$ when all vertices are disconnected(isolated) from each other, there is only one configuration of the graph, giving $\Omega=1$ and $S_{Gibbs}=0$. The same is true for the complete graph, when all vertices are connected with each other which results in $m=\frac{n(n-1)}{2}$,$R=\frac{n-1}{2}$, $\Omega=1$ and $S_{Gibbs}=0$. For the intermediate regime $0 < R < \frac{n-1}{2}$, the number of configurations $\Omega$ and the entropy Eq. \ref{entropy_gibbs} peak at some value of $R$.
This value can be derived by setting the derivative of Eq. \ref{entropy_gibbs} to zero in respect to R
\begin{gather}
\nonumber\\[1ex]
\frac{dS_{Gibbs}(n,R)}{dR}   =  0  \Rightarrow
\nonumber\\
n \left(\frac{\Gamma^{'}(1 + 1/2 (-1 + n) n - n R)}{\Gamma(1 + 1/2 (-1 + n) n - n R)}- \frac{\Gamma^{'}(1 + nR)}{\Gamma(1 +nR)} \right)= 0
\end{gather}
which gives the value of R where the entropy is maximized,
\begin{equation}
      R =  \frac{(n-1)}{4}.
\end{equation}

Now assume for example a physical quantity, or an emergent geometrical property expressed as a functional $f_i(V_j)$ of a set of $p$ vertices $V_{j}=\big\{v_{1}^{j},v_{2}^{j},..,v_{p}^{j} \big\}$ belonging to the uniform random graph $G(V,E)$. The functional $f_i(V_j)$ represents essentially a physical quantity for a subgraph $V_{j}$ of the full uniform random graph. 
Then we can formulate a physical theory on the graph, 
by defining the action 
\begin{equation}
S_{gr}=\sum_{i=1}^{N_1} \sum_{j=1}^{N_2} a_{ij} f_i(V_j)
\label{action}
\end{equation}
where $a_{ij}$ are constants. The sums run over $N_1$ different physical quantities(functionals) $f_i$ for $N_2$ different subgraphs $V_j$. By setting Eq. \ref{action} equal to a constant value ($ S_{gr} =a$) we can also formulate a constraint equation for different physical quantities on subgraph structures. In general the functional $f_i(V_j)$ represents a transformation of a set of vertices belonging to the graph. The result of the transformation can be for example another set of vertices, a scalar value of a physical quantity or an emergent geometrical property like the distance for a path between two vertices in the graph, or the scalar curvature around a vertex. This way the functional can be used to define a local or a global(topological) property for a set of vertices belonging to the graph. In this approach all physical quantities are different manifestations of local or global structural properties of the graph. In addition, equations of motion for a physical theory formulated on the graph, can be obtained by minimizing the action given by Eq.\ref{action} ($\delta S_{gr} =0$).

For example if the set for $V_j$ contains only two vertices $V_j=\big\{ v_k,v_l \big\}$, then we can define one functional $f_i(v_k,v_l)=d(v_k,v_l)$ as the distance for an arbitrary path between the two vertices. The action Eq. \ref{action}
becomes simply $S_{gr}=d(v_k,v_l)$ whose  minimization wields the distance for the path between the two vertices that consists of the minimum number of edges, that is, the geodesic distance  $d^{g}(v_k,v_l)$ between the two vertices. Therefore we have,
$S_{gr}=\min(d(v_k,v_l))=d^{g}(v_k,v_l)$.
In general, we can always split the set $V_j$ into p subsets as $V_j=\big\{ V_1,V_2,...,V_p \big\}$. Then if there are no edges between the vertices contained in different subsets, that is the subsets are disconnected and additionally if $f_i(V_j)$ represents a local quantity then we can split the functional
as
\begin{equation}
 f_i(\big\{ V_1,V_2,...,V_p \big\})=f_i(V_1) \otimes f_i(V_2) ....\otimes f_i(V_p).
\label{equation_test}
\end{equation}
This is analogous to separable states in quantum mechanics, belonging to a composite space that can be factored into individual states belonging to separate subspaces.
For example if $f_i$ represents the mass for a set of vertices
then the total mass of the set is simply the sum of the mass of the individual vertices contained in the set $f_m(\big\{ v_1,v_2,...,v_p \big\})=f_m(v_1) + f_m(v_2) ....+ f_m(v_p)$.
In this case each subgraph in Eq. \ref{equation_test} contains only one vertex.

Assuming that the number of edges m in the graph represents the total energy of the system, we can define a local energy at each vertex $v_i$ in the graph, as $E_{loc}=d(i)/2$, so that $\sum_{i=1}^{i=n}E_{loc}=m$. In the language of Eq. \ref{action}
the local energy can be represented by a functional $f_E(v_i)=d(i)/2$, with the indexes in the sums taking the values $N_1=1$, $N_2=n$ and $a_{ij}=1$, giving $S_{gr}=m$.

\section{Evolution mechanism}
In order to implement an evolution mechanism for the physical system described the graph we introduce a toy model by removing edges from an initially complete graph. We start with a complete graph at universal time $t=0$. A complete graph has all its n vertices connected with each other, giving $n(n-1)/2$ edges in total. We assume that the evolution of this physical system is achieved by removing one edge at each time tick, whose duration is represented by a fundamental quantum of time $t_f$. Essentially we consider the flow of time measured by a universal clock, whose subsequent ticks are represented by $t_f$, where one edge is removed from the graph. Therefore we assume that time is quantized in our model in fundamental time quanta $t_f$, and is represented by an external update rule for the number of edges in the graph. All time intervals are expressed as $\Delta t=i t_f$ for integer $i=0,1,2,....$.
At each step of the graph evolution, the physical system is described by one of the possible configurations $\Omega=\binom{ \binom{n}{2}}{m}$ of the m edges among the n vertices. At each step one of these configurations is picked randomly with the same probability to represent the physical state of the system.
This evolution process leads to a monotonic decrease of the Gibbs entropy Eq. \ref{entropy_gibbs}, after its maximum value is reached at $R=\frac{n-1}{4}$. This monotonic behavior of the entropy can be used as a statistical definition of a universal arrow of time for the evolving random graph.
Additionally we expect that the physical properties of the graph such as its dimensionality $D$ and other emergent geometrical properties, evolve with $t$, as the graphs anneals by removing edges. In addition, we expect that at the thermodynamic limit, for a large number of vertices ($n \rightarrow \infty$) the emergent geometrical properties of the graph should depend only on the ratio of edges over vertices $R=m/n$, i.e. the spatial density of the graph. After k evolution steps from $t=0$, corresponding to a time interval $\Delta t= k t_f$, the system will have reached the value of the ratio
\begin{equation}
R=\frac{\frac{n(n-1)}{2}-k}{n} \Rightarrow  n^2-n(1+2R)-2k=0.
\label{equation3}
\end{equation}
Solving the above equation we get the value of n
\begin{equation}
n=\frac{(1+2R)+\sqrt{((1+2R)+8k}}{2}.
\label{equation4}
\end{equation}
Assuming that $k \gg R$ after the system has evolved
sufficiently far from $t=0$ then we have
\begin{equation}
n \approx \sqrt{2k}.
\label{equation41}
\end{equation}

We have calculated numerically the radius of the graph as it
evolves by the mechanism described above. The radius $\alpha$ of a graph can be defined as the minimum, over all vertices, of the eccentricity of that vertex. That is, for x and y being two vertices of the graph $G(V,E)$ its radius is defined as
\begin{equation}
\alpha=\min_{x \in V(G)}(\max_{y \in V(G)}d(x,y))
\label{radius}
\end{equation}
where $d(x,y)$ is the distance for an arbitrary path between x and y, determined by the number of edges in the path. The definition of the radius Eq. \ref{radius} gives an estimation of the spatial extent of the graph.

\begin{figure}
\begin{center}
\includegraphics[width=0.9\columnwidth,clip=true]{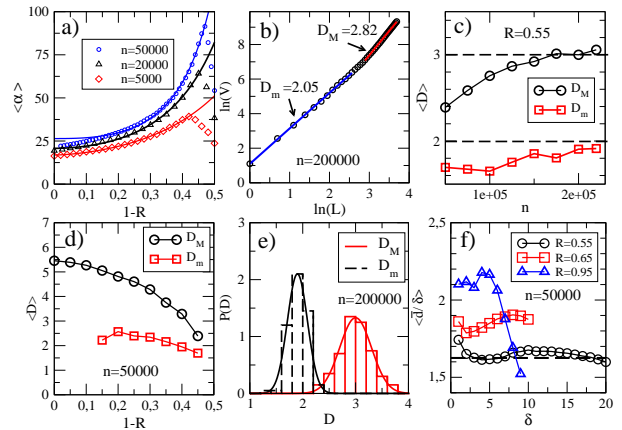}
\end{center}
\caption{a) The growth of the radius $\alpha$ of the uniform random graph as it evolves by removing one edge per a fundamental quantum of time. All curves follow an exponential growth of the form $\alpha(t_1)=ae^{b(10t_{1})^{c}}$ with $t_{1}=1-R$ with three fitting parameters $a,b,c$ with $c>1$. b) The scaling of the logarithm of the number of vertices $ln(\textrm{V})$ versus the logarithm of the linear length scale $ln(L)$, representing the different layers in the procedure for calculating the dimension of the graph. We get two different slopes for small and large $L$, by fitting with the formula $ln(\textrm{V})=\delta+Dln(L)$,
representing the microscopic $D_m$ and the macroscopic $D_M$ dimension of the graph. c)The average dimensions of the graph versus its the number of vertices $n$ at ratio of edges over vertices $R=0.55$ for 100 runs. The dimensions converge to the values $D_M \approx 3$ and $D_m \approx 2$ for large $n$. d) The evolution of average dimensions as edges are removed from the graph via the evolution mechanism. e) The probability distributions of D. The two dimensions are distributed normally, i.e. both the microscopic and macroscopic dimensional fluctuations follow the Gaussian form. f) The average distance between the surfaces of two spheres, over the distance between the sphere centers, $\frac{\overline{d}}{\delta}$ versus $\delta$. We have considered the average value of this quantity for 100 runs. The black dashed curve at the constant value $\frac{\overline{d}}{\delta }=1.625$, corresponds to a flat 3D manifold. This value is reached by the graph for ratio $R=0.55$ at large $\delta$. This result in conjunction with the macroscopic dimension reaching the value $D_M \approx 3$ in panel c), show that the graph converges approximately to a flat 3D manifold at $R=0.55$. }
\label{fig1}
\end{figure}

In the region of the ratio $R>0.5$ the graph consists of a giant
component(subgraph) and many small tree-like components that are isolated(disconnected) from the rest of the graph\cite{newman,frieze,bollob}.
We have found an exponential growth of the radius $\alpha$ of the giant component,
as the graph evolves by removing one edge per fundamental time quantum $t_f$. The result for different number of vertices $n$,
for the average radius $\langle a \rangle$ over 100 runs of the graph, can be seen in Fig. \ref{fig1}a. In all cases the radius reaches a maximum value near the critical value of the ratio $R=0.5$. At this ratio a phase transition occurs and the graph splits in many small disconnected components with a different number of vertices each\cite{newman,frieze} for $R \leq 0.5$. In the opposite region $R>0.5$, there is always a giant graph component whose radius follows the exponential growth shown in Fig. \ref{fig1}a. The maximum radius reached near $R=0.5$, increases for larger graphs corresponding to larger $n$. We expect that as the graph becomes larger, approaching the limit $n \rightarrow \infty$, the maximum radius occurs asymptotically at $R=0.5$. The exponential growth follows the form $\alpha(t_1)=ae^{b(10t_{1})^{c}}$ with $t_{1}=1-R$, with three fitting parameters $a,b,c$, represented by the fitting continuous curves in Fig. \ref{fig1}a. The fitting parameters determine the rate of expansion. We have found that $c>1$ for all the $n$ studied. In addition, we have found the same exponential growth for the geodesic distance $d^{g}(v_j,v_k)$ between two arbitrary vertices $v_j,v_k$ in the graph. From Eq. \ref{equation4}, since $k \approx n^2$, where $k$ is the number of evolution steps, we can see that larger graphs need more time to reach the same ratio $R$ and obtain the relevant geometrical properties.

\section{Dimension}
One of the central concepts characterizing a metric space is its spatial dimension. The usual definition of the dimension counts the minimum number of coordinates in an orthogonal coordinate system needed to describe all the points of the metric space.
We examine/address the fundamental question whether a random discrete structure like a uniform random graph $G$ can converge to a  continuous metric space(D-dimensional manifold $M$) at the limit of a large number of vertices $G \rightarrow M$. We do that by calculating the spatial dimension of the space emerging by the connectivity properties of the graph. We follow a scaling approach by examining how the connected vertices form an ambient space, in which the graph can be embedded so that it contains no crossing edges. This method is similar to the calculation of the fractal/Hausdorff dimension but without pre-embedding the graph in an ambient space with a fixed coordinated system. Instead the dimension in our method is solely determined by the connections between the vertices, and the metric space can emerge gradually as additional layers in the scaling are considered.
We define a layer, $L_{G}^{r}(v_{i})$, of radius $r$ around a vertex $v_{i}$ belonging to the graph $G$, as the set of vertices
\begin{equation}
L_{G}^{r}(v_{i}) =	\{ v \in G | d^{g}(v_{i},v) =r, r\geq 1 	\},
\end{equation}
where $d^{g}(v_{i},v)$ is the minimum distance (geodesic) between vertices $v_{i}$ and $v$. Using the language of Eq. \ref{action} we can define a functional $f_{r}(v_{i})$ that acts on vertex $v_{i}$ as
\begin{equation}
 f_{r}(v_{i})=L_{G}^{r}(v_{i}) \setminus L_{G}^{r-1}(v_{i})  \setminus... \setminus L_{G}^{1}(v_{i}),
\end{equation}
ensuring that a layer at geodesic distance $r$ does not contain vertices that have been already included in the previous layers $r-1,...,1$. The number of vertices after $\textrm{L}$ layers is,
\begin{equation}
\sum_{r=1}^{\textrm{L}} |f_{r}(v_{i})|=  \textrm{V}(\textrm{L}), 
\end{equation}
which is equivalent to the action Eq. \ref{action}, where the sum over functionals runs over the different layers ($N_1=L$) and the sum over subgraphs contains only vertex $v_i$ ($N_2=1$). The term
$\textrm{V}(\textrm{L})$ counts the number of vertices contained in a ball of radius L around vertex $v_i$,
\begin{equation}
B_{G}^{\textrm{L}}(v_{i}) =	\{ v \in G | d^{g}(v_{i},v) =r, r\leq \textrm{L} \},
\end{equation}
that is $\textrm{V}(\textrm{L})=|B_{G}^{\textrm{L}}(v_{i})|$.
If $\textrm{V}(\textrm{L})$ grows with L as
\begin{equation}
\textrm{V}(\textrm{L}) \sim  \textrm{L}^{D}, 
\label{scaling_dimension}
\end{equation}
that is if the graph locally at each vertex follows the ball topology, then we define the exponent D as the spatial dimension of the graph. Essentially Eq. \ref{scaling_dimension} counts the number of vertices contained in a geodesic ball  of radius $\textrm{L}$ and volume $\textrm{L}^{D}$. When $D$ is an integer then the graph can be embedded into a D-dimensional manifold without any crossing edges. In this case all the vertices can be described approximately by an orthogonal coordinate system with D spatial directions (D independent variables) and the graph resembles a D-manifold. The resemblance/convergence to the D-manifold for integer values of $D$ is further supported by curvature calculations that we present also in the current manuscript. 

We have calculated the spatial dimension for the giant graph component, in the region $R>0.5$ via the scaling approach presented above. We have performed the calculation by using the center of the graph component as the origin of the scaling, in order to avoid boundary effects at the periphery of the graph, and also considered 100 runs of the random graph. For a fixed ratio $R$ the value of the dimension will fluctuate among the different runs. Since the random graph is an ergodic system this method to study the statistical properties of $D$, like its fluctuations and its average value, will be equivalent to using different vertices in the graph as origins of the scaling method used to calculate $D$, as long as the periphery of the graph is avoided. In Fig. \ref{fig1}b we show an example for the scaling of the logarithm of the number of vertices $\ln{\textrm{V}}$ versus the logarithm of the scaling length $\ln{\textrm{L}}$ for a graph with $n=200000$ at ratio $R=0.55$ for one run of the graph. There are two distinct scaling behaviors of the type Eq. \ref{scaling_dimension}, for small and large $\textrm{L}$, described by two different slopes, represented by the two linear fitting curves of different color in Fig \ref{fig1}b. For this reason we define two dimensions, a microscopic one $D_{m}$ for small $\textrm{L}$ and a macroscopic one $D_{M}$ for large $\textrm{L}$. We have observed similar linear growth behaviors for other $R$, which allows us to calculate the dimension of the random graph. The linear growth becomes more clear with reduced statistical fluctuations as $R$ increases. The almost perfect linear growth of $\ln{V}$ with $\ln{\textrm{L}}$ for the two different scaling regions shown in Fig \ref{fig1}b, implies that the graph obtains a fractal structure near $R=0.5$. This is consistent with the phase transition occuring at this critical $R$ leading to the appearance of many small disconnected components(subgraphs) for $R \leq 0.5$, instead of a giant one and many tree-like ones for $R>0.5$\cite{newman,frieze,bollob}.
In addition we note that the small tree-like graph components that appear in the region $R>0.5$ alongside the giant component, whose dimension we have studied, can be thought as small isolated spaces of low dimensionality.

As shown in Fig. \ref{fig1}c the average macroscopic dimension reaches the value $D_{M} \approx 3$ at $R=0.55$ for a large number of vertices n, at the end of the exponential expansion of the graph, where the maximum radius is reached in Fig. \ref{fig1}a. In addition the average microscopic dimension in Fig. \ref{fig1}c reaches the value $D_{m} \approx 2$ for large n as in other discrete model approaches for spacetime\cite{markopoulou,loll2}. Our result shows that a uniform random graph modeling of space leads to a random structure that resembles macroscopically a 3D manifold, and microscopically a 2D manifold.  In Fig. \ref{fig1}d we show the evolution of $D_M$ with $R$ for $n=50000$. The dimensionality of the random graph evolves as it expands, implying an evolving dimensionality of the emergent space. We note that in one of our previous works\cite{paper1} we have derived $D \approx 2$, for small graphs at $R=0.67$, corresponding to the microscopic dimension in the current paper. In this paper we make also an analogy to 2D honeycomb lattices(graphene), and compare the electronic properties of the two systems, implying possible relativistic effects for the small random graphs, corresponding to small scales $\textrm{L}$ in the current model. We note also that in our model the state of space at $t=0$ is represented by a complete graph which has infinite dimension,
since all vertices are connected with each other. This could represent the big-bang singularity in the evolving cosmological models of the universe, such as cosmic inflation.

In Fig. \ref{fig1}e we show the fluctuations of the dimension, via the respective probability distributions for 100 runs of the random graph, at $R=0.55$ and $n=200000$.
Both the micro and macro dimensions are distributed normally
around the values $D_m=2$ and $D_M=3$.
As the graph evolves, due to ergodicity, these dimensional fluctuations will manifest along the different vertices in the graph, which correspond to different spatial positions in the emergent space. The fluctuations should follow the normal distribution that we have shown, for small time scales. A fundamental question that arises is whether these dimensional fluctuations can manifest physically as extra degrees of freedom for emergent particles propagating inside the graph, which will act as a vacuum with the respective average dimensionality $D_m=2$ for microscopic and $D_M=3$ for macroscopic scales.

\section{Curvature}
In order to study the curvature generated by the uniform random graphs, we use a general method formulated in reference \cite{klitgaard,klitgaard1}, which focuses on regular/periodic lattices and triangulated surfaces in two dimensions (2D). The method is also valid for any discrete structure, like the uniform random graphs that we study in the current paper. The quantum Ricci curvature between two vertices $K(p,p')$ in the graph can be defined by examining how the average distance $\overline{d}$ between the surfaces of two spheres centered at vertices $p$ and $p'$, changes with the distance between the two vertices $\delta$. This definition essentially expresses how the curvature at vertex $p$ changes as we move in the direction of vertex $p'$. The quantum curvature converges to the regular Ricci scalar curvature for continuous manifolds for large $\delta$ as shown in\cite{klitgaard}. In order to extract the quantum curvature we calculate the following quantity
\begin{gather}
\overline{d}( S^{\epsilon}(p), S^{\epsilon}(p') )   =   
\nonumber\\
\frac{1}{N_{0}( S^{\epsilon}(p))} \frac{1}{N_{0}( S^{\epsilon}(p'))} \sum_{q\in S^{\epsilon}(p)} \sum_{q'\in S^{\epsilon}(p')} d(q,q'),
\label{discrete}
\end{gather}
where we consider two spheres centered at vertices $p$ and $p'$, of radius $\epsilon$, which is equal to the geodesic distance $\delta$ between $p$ and $p'$, that is $\epsilon=\delta$. Also, $d(q,q')$ is the geodesic distance between vertices $q$ and $q'$ which lie at the surfaces of the two spheres and $N_{0}( S^{\epsilon}(p))$ ,$N_{0}( S^{\epsilon}(p'))$ are the corresponding number of vertices lying at the surfaces. The scaling behavior of Eq. \ref{discrete} with $\delta$ allows us to extract the quantum Ricci curvature $K(p,p')$ via
\begin{equation}
 \frac{\overline{d}( S^{\delta}(p), S^{\delta}(p') )}{\delta }= c(1-K(p,p')),
 \label{discrete1}
\end{equation}
where $c$ is a positive constant that depends on the metric structure of the simplicial manifold under consideration. We have performed the calculation for vertices near the center of the graph in order to avoid finite size effects, induced by the periphery of the graph. In Fig. \ref{fig1}f we plot the average value of Eq. \ref{discrete1} for 100 runs of the graph and for different ratios $R$ for $n=50000$. Also, we have included the value of Eq. \ref{discrete1} for flat continuous 3D manifolds $\frac{\overline{d}}{\delta }=1.625$, which is constant not depending on $\delta$, represented by the black dashed line in Fig. \ref{fig1}f. For $R=0.95$ at large $\delta$, the graph has a positive curvature since Eq. \ref{discrete1} decreases with $\delta$. As R is reduced and the graph becomes less dense, the slope of Eq. \ref{discrete1} decreases and therefore the corresponding curvature becomes less positive. At $R=0.55$ the value of Eq. \ref{discrete1} coincides with that of the flat 3D manifold(black dashed line). Therefore the graph is almost flat with a slightly positive curvature at $R=0.55$. This result along with the dimensionality reaching the value $D_M \approx 3$ for large $n$, suggests that the uniform random graph geometrically reduces approximately to a flat 3D manifold at ratio $R=0.55$.

In addition, it has been shown that any random graph with degree $d(i)<6$ can be embedded into a 3D manifold/space without any crossing edges\cite{cohen}, which is the case for the uniform random graph at $R=0.5$, being sparse with a low degree at each vertex. The embedding theorem along with the graph dimension reaching the value $D_{M}=3$ and the curvature the value $K=0$ at $R=0.5$, hint that the graph at large scales converges to a random structure consisting of tangent 3D polyhedra/polytopes with a varying number of flat polygonal faces. At small scales, since we have found $D_{m}=2$ the graph becomes planar instead resembling a 2D manifold.

Finally we note that, since both the dimension and the curvature emerge simultaneously out of the connectivity properties of the random graph, their contribution to the geometrical properties can be mixed\cite{klitgaard1}. For example a graph with a non-integer(fractional) dimension D can resemble a manifold with integer value of the dimension closest to D, with a non-zero curvature. For instance the graph at $R=0.95$ where $D_{M} \approx 5.4$, resembles a five-dimensional manifold with a positive curvature.

\section{Inflation}
We can use Eq. \ref{equation4} to estimate the number
of the vertices in the graph needed to reproduce the
times conjectured in cosmic inflation models. The inflationary epoch, is conjectured to have lasted from $10^{-36}$ seconds after the Big Bang singularity, represented by a complete graph in our model, to some time between $10^{-33}$ and $10^{-32}$ seconds after the singularity. In addition, we can assume that our model describes the exponential expansion of space until it reaches the dimensionality D=3, corresponding to ratio $R=0.5$. We have assumed that every tick in our model lasts a fundamental quantum of time $t_f$. Assuming that this represents the Planck time $t_f=t_p=10^{-44}s$, we need $k=10^{12}$ ticks to reach the  time $10^{-32}$ seconds, where the inflationary epoch ends. Then from Eq. \ref{equation4} we get the number of vertices $n \approx 10^6$ in the graph, needed to reproduce the times of the inflationary epoch.

\section{Relativistic effects}
An Einstein-Hilbert-like action can be defined
by using Eq. \ref{action}. We consider a functional
that represents the scalar curvature $f_{i}(v_j)=\textrm{k}(v_j)$
for a vertex $v_j$, defined as the sum of the quantum Ricci curvature $K(v_j,v_k)$ over its neighboring vertices $v_k$ as $\textrm{k}(v_j)=c_d \sum_{v_k}^{d(i)} K(v_j,v_k)$, where $K$ is given by Eq. \ref{discrete1}, $d(i)$ is the degree of $v_j$ and $c_d$ is a factor depending on the local dimension of the graph. In addition by considering the analogy to the  Einstein-Hilbert-like action we can define the rest of the constants in Eq. \ref{action} as $a_{ij}=c_{ij} \sqrt{g}$, where $g$ is the determinant of the metric $g=det{g_{\mu \nu}}$ for the vertex $v_j$ and $c_{ij}$ is a normalization factor.
Any graph can be considered locally flat at a vertex
and its nearest neighboring area, so that the local metric
$g_{gr}$ for any vertex in the graph is a unit matrix.
Each edge connecting the vertex to its neighboring vertices can be considered as a spatial direction(axis) for a local orthogonal coordinate system of dimension $D=d(i)$ defined at the vertex.  Therefore the local metric at this vertex is a unit matrix of size $d(i) \times d(i)$ resulting in $g=det(g_{\mu \nu})=1$ and $\sqrt{g}=1$. Note that the respective term in the Einstein-Hilbert action has the form $\sqrt{-g}$, in order to ensure that it takes the value one, when spacetime is flat and the metric has the property $g=det(g_{\mu \nu})=-1$. In our case we assume that the metric contains only the space part, as time is simulated via an external update rule for the number of edges in the graph. Then the sum in the action for the graph Eq. \ref{action}, running over the different subgraphs, becomes simply the sum of the scalar curvature $\textrm{k}(v_j)$ over all the vertices of the random graph $(N_2=n)$. In order to implement the integration over time also, as in the Einstein-Hilbert action, we can consider the evolution mechanism of the graph.
As each step of the evolution, where one edge is removed from the graph, the physical system is described by one of the possible configurations $\Omega=\binom{ \binom{n}{2}}{m}$ of the m edges among the n vertices, which is picked randomly with the same probability to represent the physical state of the system. Then the sum over different functionals in Eq. \ref{action} becomes the sum over the scalar curvatures $\textrm{k}(v_j)$ for each step in the evolution of the graph where $N_1=k$ determines the number of evolution steps, corresponding to time interval $\Delta t=k t_f$. Then the action of the graph Eq. \ref{action} can be written as,
\begin{equation}
S_{gr}= \sum_{i=1}^{k} \sum_{j=1}^{n} c_{ij} \textrm{k}^{n,m-i}(v_j).
\label{equation1}
\end{equation}
 The above action, can be minimized in respect to the different possible paths of the evolution of the graph, in order to derive classical equations of motion for the graph. This process is analogous to the minimization of the Einstein-Hilbert action in respect to the metric, that leads to the Einstein-field-equations(EFE).

A relative notion of time can emerge from our model by considering the rate of change between two physical quantities/functionals on different subgraphs as defined above. For example we can define an operator
\begin{equation}
\partial^{ij}_{kl} =\frac{\partial f_i({V_j})}{\partial f_k({V_l})},
\label{equation5}
\end{equation}
which expresses the change of a physical quantity $f_i$ on subgraph
${V_j}$ relative to a physical quantity $f_k$ on subgraph
${V_l}$. This way the notion of time dilation from relativity can emerge from our model. For example regions of the graph with different curvatures could evolve at different rates, higher curvature corresponding to slower evolution rate. In this approach means of measuring time like clocks are defined using the evolution of subgraph structures.

In addition it has been shown that the Lorentz invariance
can be formulated on discrete models like periodic lattices
in various dimensions\cite{disclor}.

\section{Mass-energy-space equivalence}
In our model all physical properties arise
as different manifestations of the connectivity properties
of the graph, including space and matter-energy. By taking account of this fundamental principle a mass-energy-space equivalence naturally arises from our model. For example we can assign a nominal mass to each vertex in the graph, expressed by a functional $f_{1}(v_j)=m_{v}$ in Eq. \ref{action}, which satisfies the property Eq. \ref{equation_test}, that is, the total mass of a subgraph is simply the sum of the masses of the individual vertices contained in the subgraph. In addition we can choose as a subraph the set of vertices contained in a geodesic ball of radius L around vertex $v_j$, which is determined by a scaling relation of the type $\textrm{V}=\epsilon+\zeta \textrm{L}^D$ (Eq. \ref{scaling_dimension}). The total mass contained in the ball of volume $W=(\textrm{L}l_f)^D$ is $M=m_{v}\textrm{V}$,
where $l_f$ is a fundamental quantum of length, representing for example the edge length in the graph. The density of vertices in the graph is
\begin{equation}
\rho_{gr}= \frac{M}{W}=\frac{m_v}{l_{f}^{D}}( \, \frac{\epsilon}{\textrm{L}^D}+\zeta) \,.
\label{emse}
\end{equation}
Assuming that $L \gg 1$ we have
\begin{equation}
\rho_{gr}= \frac{\zeta m_v}{l_{f}^{D}}.
\label{emse1}
\end{equation}
For $D=3$ the above formula corresponds to the energy density of the vacuum $\rho_{vacuum}=\rho_{gr}$, which is related to the cosmological constant $\Lambda$ in EFE via $\Lambda=\kappa \rho_{vacuum}$ where $\kappa=\frac{8 \pi G}{c^4}$ is Einstein's gravitational constant. From this perspective the cosmological constant can be thought as an intrinsic property of space as it emerges from the graph structure. Both $D$ and $\zeta$ will fluctuate as we have shown in Fig. \ref{fig1}e and consequently the density $\rho_{gr}$ will also fluctuate. Using the language of Eq. \ref{action} we can define two functionals for the subgraph of vertices contained in a geodesic ball of radius L around vertex $v_j$, one for the mass $f_1(v_j)=M=m_{v}V$ and one for the volume $f_2(v_j)=W=(\textrm{L}l_f)^D$. By setting Eq. \ref{action} equal to zero we can write,
\begin{gather}
\begin{aligned} 
\nonumber a_{11} f_1(v_j)+a_{21} f_2(v_j)=0 \Rightarrow a_{11} M+a_{21} W=0 &  \\
\nonumber \Rightarrow M=-\frac{a_{21}}{a_{11}} W 
\Rightarrow M= \rho_{gr} W \Rightarrow  M= \rho_{vacuum} W \Rightarrow 
\end{aligned}\\
\nonumber M= \frac{\Lambda}{\kappa} W
\end{gather}
\label{equation2}
The above formula can interpreted as a way to transform between a volume of empty space $W$, and matter-energy $M$, via the energy-density of the vacuum or the cosmological constant. Note that this is different from being able to extract energy from the vacuum that has been proposed in various studies. Therefore, in our uniform random graph approach a unification of space and matter-energy, and a way to transform between them via the cosmological constant emerges naturally. The unification can be considered as a natural extension of the mass-energy equivalence $E=mc^2$ in relativity, expressed as a mass-energy-space equivalence.
Our approach suggests that the cosmological constant does not originate from quantum fluctuations of fields that fill empty space. It can be thought instead as an intrinsic property of space as it emerges from random discrete structures. like the uniform random graphs considered in the current paper. Our approach could provide an explanation of the low value of the cosmological constant measured experimentally.
Interestingly the mass-energy-space equivalence is already hinted in GR. For example the cosmological term $\Lambda g_{\mu \nu}$ can be moved freely on the left or the right side of the EFE, interpreted
either as an intrinsic curvature of empty space or as a part of the stress-energy tensor $T_{\mu \nu}$, which represents the effects of matter-energy. If we consider it as part of the stress-energy tensor then we have
\begin{equation}
T^{vac}_{\mu \nu}= - \frac{\Lambda}{\kappa} g_{\mu \nu}.
\label{efe_equi1}
\end{equation}
The element $T^{vac}_{00}=\frac{E}{W c^2}=\frac{M}{W}$ is the energy per unit volume divided by $c^2$ (mass density) resulting in $M= \frac{\Lambda}{\kappa} W$, since $g_{00}=-1$.
\section{Uncertainty principles} 
We can define generalized uncertainty principles, by using the statistical fluctuations of different physical quantities defined on the graph. For example we can use the product of the variances $\Delta A \Delta B$, where A and B are two different physical quantities, in analogy to the Heisenberg uncertainty principle. Consider the local energy $E_{loc}=d(i)/2$ at each vertex, which will fluctuate, following the hypergeometric distribution, with variance
\begin{equation}
\Delta(E_{loc})=\frac{\Delta(d(n,m))}{2}\frac{\hbar}{t_f} =\frac{2m(n^{2} -2m -n)}{2n^{2}(1+n)} \frac{\hbar}{t_f},
\label{hypervar}
\end{equation}
 where $E_f=\frac{\hbar}{t_f}$ is a fundamental unit of energy in our model, expressed in units of fundamental quantum of time $t_f$, which determines the duration of each step in the discrete evolution mechanism of the graph. Note that the variance from the above formula decreases with decreasing $m$. For example at the limit $n \rightarrow \infty$ we have $\Delta(E_{loc})=2$ for $m=2n, R=2$ and $\Delta(E_{loc})=1/2$ for $m=n/2, R=0.5$. As each step of the graph evolution, where one edge is removed from the graph, we assume that the physical system is described by one of the possible configurations $\Omega=\binom{ \binom{n}{2}}{m}$ of the m edges among the n vertices, all having the same probability to appear $1/\Omega$. All time intervals and macroscopic energies along with uncertainties(statistical fluctuations) can be expressed in units of the two fundamental constants as $\Delta t=i t_f$ and $\Delta E=j E_f$ where $i,j$ are integers $i=0,1,2,....,j=0,1,2,....$.
 In a time interval $\Delta t=k t_f$, where k is the number of time steps in the graph evolution, the fluctuations of the $E_{loc}$ will consist of the fluctuations of k independent statistical ensembles with a different number of edges determined by the evolution mechanism. Based on this idea, and using Eq. \ref{hypervar}, the variance of the local energy $\Delta E$ in a time interval $\Delta t$ can be written as
 \begin{equation}
 \Delta E=\Delta(\sum_{i=0}^{k} E_{loc}(i)) =\frac{\sum_{i=0}^{k}\Delta(d(n,m-i))}{2k^2}\frac{\hbar}{t_f}.
\label{elocvar}
\end{equation}
where $k \leq m$. From the above equation we can see that the variance(uncertainty) of the local energy $\Delta E$ is inversely proportional to k, which determines the duration of the time interval $\Delta t=k t_f$, in which the measure of the energy is performed, as in the  Heisenberg uncertainty principle. We can write
\begin{equation}
 \Delta E \Delta t =\frac{\hbar \sum_{i=0}^{k}\Delta(d(n,m-i))}{2k}.
 \label{test1}
\end{equation}
By using Eq. \ref{hypervar} the sum(series) in the above equation, excluding the factor $\frac{\hbar}{2}$, can be written as
\begin{equation}
   -\tfrac{(1+k)(4 k^2 + k (2 - 12 m + 3 (-1 + n) n) +  6 m (2 m + n - n^2))}{3kn^{2} (1 + n)} 
\end{equation}
By taking the thermodynamic limit $n \rightarrow \infty$ and considering that $m=Rn$, the above formula satisfies the condition
\begin{equation}
2 R \left( \frac{1}{k} + 1 \right) \geq 1, 
\end{equation}
for $R \ge 0.5$. Therefore at the limit $n \rightarrow \infty$ we have
\begin{equation}
\frac{\sum_{i=0}^{k}\Delta(d(n,m-i))}{k} \geq 1,
\label{test2}
\end{equation}
which transforms Eq. \ref{test1} to
\begin{equation}
\Delta E \Delta t \geq \frac{\hbar}{2}. 
\end{equation}
Consequently, the Heisenberg uncertainty principle emerges naturally from the statistical fluctuations of the degrees in our graph model.

We can replace the fluctuations(variance) of the degree $\Delta(d(n,m-i))$ in Eq. \ref{test1} with the variance of the dimension $\Delta D$, which follows the Gaussian distribution as we have shown in Fig. \ref{fig1}e. Then if Eq. \ref{test2} is satisfied we end up with an uncertainty principle for the dimension and time $\Delta D \Delta t \geq constant$, implying that the dimensional fluctuations could be detectable in high-energy experiments involving short time scales.

The dimension and the volume are examples of two commutable physical quantities in our model, that do not satisfy an uncertainty principle. Since the dimension D fluctuates normally with variance $\Delta D$, the corresponding volume $W=\textrm{L}^D$ (for $l_f=1$) will fluctuate log-normally with variance given from the following formula $\Delta W=(\exp^{\Delta D (\ln \textrm{L})^{2}}-1)\exp^{\left(2 \langle D \rangle \ln \textrm{L} + \Delta D (\ln \textrm{L})^{2} \right)}$.

\section{Emergent particles}
Finally, particles in our model can be thought as persistent subgraph structures propagating inside the graph which acts as a uniform background/vacuum for these excitations. The equations of
motion of these persistent structures should follow the Lorentz transformation rules from relativity. For example we have shown the existence of 0D and 1D persistent localized states in a previous study of the energy/spectral and localization properties of uniform random networks\cite{paper1}. The 0D states lie at energy E=0 while the 1D states occur at various non-zero energies and are consisted of line of vertices of various lengths, resembling strings. In addition these emergent particles could be transformed to empty space via the mass-energy-space equivalence that we have proposed. In the same work we have shown a linear behavior of the density of states, leading to a linear energy dispersion $E=\hbar v_{nt}k$ relation at the band center of the system at energies E=0, as for relativistic massless particles with effective network velocity $v_{nt}=\frac{1}{\hbar}\sqrt{\frac{2}{\pi \alpha}}$, where $\alpha$ is the slope of the DOS at E=0. In our model the fluctuations of the dimension D lead to fluctuations of the volume of empty space W. By applying the mass-energy-space equivalence that we have proposed the fluctuations of the empty space can be transformed to massive particles, represented by persistent subgraph structures as described above. These particles essentially emerge indirectly out of the fluctuations of the dimension. This mechanism bears similarity to the quantum fluctuations of vacuum that give rise to virtual particles, obeying the Heisenberg uncertainty principle, in quantum field theory.

\section{Conclusions}
We have studied the geometrical properties emerging out of the connectivity properties of uniform random graphs. We have shown that graphs with ratio of edges over vertices $R=0.5$
have average spatial dimension $D=3$ and curvature $K=0$, implying  a convergence to a flat 3D manifold. These properties coincide with the appearance of the large component of the graph at $R=0.5$. In addition we have introduced an evolution mechanism for the graph by removing one edge per a fundamental quantum of time. We have shown that this mechanism leads to an exponential growth of the radius of the graph, reaching a maximum at $R=0.5$, bearing similarities to the exponential expansion of space in cosmic inflation models. We have introduced a generalized physical action for graphs based on functionals over different subgraph structures. By using this action we have shown how general relativistic effects can be recovered via an Einstein-Hilbert-like action, using the scalar curvature defined around each vertex. Additionally we have proposed a mass-energy-space equivalence, arising naturally from our graph model approach, that leads to a way to transform between empty space and matter-energy via the cosmological constant. Finally we have shown the emergence of the energy-time Heisenberg uncertainty principle from the statistical fluctuations of the degrees in the graph. In conclusion, our results show the possibility of 3D flat space, its geometrical, structural and quantum mechanical properties along with its interaction with matter-energy, all emerging from a uniform random graph approach.

\section{acknowledgements}
We acknowledge support by the Project HPC-EUROPA3, funded by the EC Research Innovation Action under the H2020 Programme. In particular, we gratefully acknowledge the computer resources and technical support provided by ARIS-GRNET and the hospitality of the Department of Physics at the University of Ioannina in Greece.

\section*{Data Availability Statement}
This manuscript has no associated data or the data will not be deposited. All the numerical data from our calculations are displayed/plotted inside the figures.

\section*{Author Contribution Statement}
Both authors I.K. and I.A. contributed equally to the design and implementation of the research, to the analysis of the results and to the writing of the manuscript.


\begin{thebibliography}{99}

\bibitem{sunil}S. Mukhi,  Class. Quantum Grav.  {\bf 28}, 153001 (2011),
\bibitem{dienes}K. R.Dienes,  Phys. Rept.  {\bf 287}, 447 [hepth/9602045] (1997),
\bibitem{rovelli1}  Class. Quantum Grav. {\bf 28}, 153002 (2011).
\bibitem{rovelli2} C. Rovelli,  Living Rev. Relativ. {\bf 11}, 5 (2008).
\bibitem{bombelli}L. Bombelli, J. Lee,  D. Meyer and R. D. Sorkin, Phys. Rev. Lett.  {\bf 59},   521-524 (1987).
\bibitem{fay1}F. Dowker, Annals of the New York Academy of Sciences {\bf 1326}, 18-25, (2014).
\bibitem{fay2}F. Dowker and S. Zalel,   C R Physique {\bf 1}, 8:246-253 (2017).
\bibitem{surya}S. Surya,  Living Rev. Rel. {\bf 22},  5 (2019), arXiv:1903.11544 [gr-qc]
\bibitem{wolfram}S. Wolfram, Complex Systems {\bf 29}, (2)   pp. 107-536 (2020).
\bibitem{gorard}J. Gorard Complex Systems, {\bf 29}, (2)  pp. 599?654 (2020).
\bibitem{markopoulou}T. Konopka, F. Markopoulou and S. Severini, Phys. Rev. D {\bf 77}, 104029 (2008).
\bibitem{lombard}J. Lombard,  Phys. Rev. D  {\bf 95},  024001
(2017).
\bibitem{trugenberger}C.A. Trugenberger, J. High Energ. Phys. , 45 (2017).
\bibitem{trugenberger2}C. Kelly, C. Trugenberger, and F. Biancalana, Class. Quantum Grav. {\bf 38}, 075008 (2021).
\bibitem{trung} Trugenberger, C.A., J. High Energ. Phys.  {\bf2022}, 19 (2022). 
\bibitem{loll1} J. Ambjørn, A. Görlich, J. Jurkiewicz, and R. Loll, y, Phys. Rep. {\bf 519}, 127 (2012)
\bibitem{loll2} R Loll 2020 Class. Quantum Grav. 37 013002
\bibitem{klitgaard} N. Klitgaard and R. Loll, Phys Rev D.{\bf 97} 046008 (2018).
\bibitem{klitgaard1} N. Klitgaard and R. Loll, Phys. Rev. D {\bf 97}, 106017 (2018)
\bibitem{farkas}I. J. Farkas, I. Der\'{e}nyi, A.-L. Barab\'{a}si, and T. Vicsek,  Phys. Rev. E {\bf 64}, 026704 (2001)
\bibitem{newman}M. E. J. Newman  Networks: An Introduction (Oxford University Press, Oxford, 2010)
\bibitem{frieze}Frieze A, Karonski M: Introduction to random graphs. Cambridge University Press (2015)
\bibitem{bollob}B. Bollob\'as, Random graphs, Cambridge University Press, Cambridge, 2nd edition, (2001).

\bibitem{cohen} R. Cohen, P. Eades, T. Lin and F. Ruskey, Lecture Notes in Computer Science 894, Springer-Verlag, 1995, 1-11

\bibitem{kosmidis}Daqing, L., Kosmidis, K., Bunde, A. et al., Nature Phys 7, 481-484 (2011).


\bibitem{paper1} I. Kleftogiannis and I. Amanatidis
Phys. Rev. E {\bf 105}, 024141 (2022).

\bibitem{disclor} Pei Wang 2018 New J. Phys. 20 023042



\end{thebibliography}
\end{document}